%% file: ms.tex
\documentclass{article}

\usepackage[utf8]{inputenc} 
\usepackage[T1]{fontenc}    
\usepackage{hyperref}       
\usepackage{url}            
\usepackage{booktabs}       
\usepackage{amsfonts}       
\usepackage{nicefrac}       
\usepackage{microtype}      
\usepackage{graphicx}
\usepackage{amssymb}
\usepackage{caption}
\usepackage{subcaption}
\usepackage{color}
\usepackage{enumitem}
\usepackage[noend]{algpseudocode}

\usepackage[toc,page]{appendix}
\usepackage{float}
\usepackage{multirow}
\usepackage{longtable}

\usepackage{dirtytalk}
\usepackage{natbib}

\newcommand{\SA}[1]{{\color{black}{#1}}}

\title{A logical set theory approach to journal subject classification analysis: intra-system irregularities and inter-system discrepancies in Web of Science and Scopus}

\author{Shir Aviv-Reuven\thanks{avivres@biu.ac.il} , Ariel Rosenfeld\thanks{ariel.rosenfeld@biu.ac.il}\\ Department of Information Sciences, Bar-Ilan University, Israel}

\begin{document}

\maketitle
\begin{abstract}
Journal classification into subject categories is an important aspect in scholarly research evaluation as well as in bibliometric analysis. 
However, this classification is not standardized, resulting in several different journal subject classification systems.
In this study, we adopt a logical set theory-based  definition of irregularities within a given classification system and discrepancies between systems and investigate their prevalence in the two most widely used indexing services of Web of Science (WoS) and Scopus.
In both systems, we identify unusually sized categories, high overlap and incohesiveness between categories. In addition, across the two systems, journals are systematically classified to a different number of categories and most categories in either system are not adequately represented in the other system. 
Our findings suggest that these \SA{irregularities} and discrepancies are, in fact, non-anecdotal and thus cannot be easily disregarded. \SA{Consequently, potentially misguided and/or inconsistent outcomes may be encountered when relying on these subject classification systems.}  
\newline
\newline
{\bf keywords:}
Journal subject classification $\cdot$ Scientometrics $\cdot$ Logical Set theory $\cdot$ Web of Science $\cdot$ Scopus

\end{abstract}
\input{Introduction}

\input{Background}

\input{Data_collection} 
\section{Results and Discussion}
\label{sec:data_analysis}

\input{Results_intra}

\input{Results_inter}

\input{conclusion.tex}
\section*{}
Declarations of interest: none
\newline
\newline
This research did not receive any specific grant from funding agencies in the public, commercial, or not-for-profit sectors.

\newpage

\bibliographystyle{apalike}
\bibliography{ms}

\end{document}

%% file: Introduction.tex
\section{Introduction}
\label{sec:intro}

Journal classification into subject categories is an  important aspect of the journal indexing systems.
From a theoretical perspective, this classification is an external expression of the internal structure of science and thus, it can foster research on the inherent relationships between scientific fields, institutes and researchers as well as many other scientometric phenomena. From a practical perspective, this classification is often used by researchers in order to find information related to their field of work and allows one to avoid sifting through possibly many irrelevant journals. Universities and other institutions also rely on these subject categories for their evaluation and, in many cases, request their researchers to publish their works in journals in a specific set of classifications or in journals which are highly ranked in their classified subject categories. \SA{For example}, researchers are mainly evaluated based on the articles they published in high ranking journals in their respective field (e.g., top 25\% of the journals in a subject category) \citep{dennis2006research, mckiernan2019meta, rice2020academic}.

Unfortunately, classifying  journals into subject categories is an ill-defined problem since the delineation of a scientific field of research is, itself, unclear and journals' boundaries need not necessarily align with those of any given field of study. Several key challenges include  interdisciplinarity, multipisciplinarity and the dynamic nature of scientific enquiry \citep{zitt2019bibliometric}. While various frameworks for the delineation have been proposed in the past (e.g., \citet{hammarfelt2017scientific}), these differ in the structure of the classification, its granularity, semantic aspects and additional parameters (see \citet{waltman2019field, archambault2011towards}). As such, major indexing services such as Web of Science (WoS), Scopus and others have developed their own unique journal subject classification systems. 

Journal classification systems use a variety of (partially) overlapping and non-exhaustive subject categories. This results in many journals being classified into more than a single category. \SA{However, the classification is not standardized and, as such, various irregularities \textit{within} a given system and discrepancies \textit{between} different systems may be encountered. 
Irregularities may misguide the  users of these systems, potentially leading to unwarranted insights and conclusions. Discrepancies, on the other hand, suggest that the outcomes from one system may be inconsistent with those derived from another.}

\SA{Logical set theory lends itself to mathematically defining such notions, as it is the main tools for analysing sets and their relationships. In our context, categories are viewed as sets of journals. As such, within a given journal subject classification system, we define three types of intra-system irregularities: 

\begin{enumerate}
    \item \textit{Size}: Categories which lay at the extremes of the category size distribution, i.e., very small or very large.
    \item \textit{Similarity}: Pairs of categories for which the relative number of shared journals (i.e., journals which are at the intersection of both categories) is large. 
    \item \textit{Coverage}: Categories in which \textit{all} journals are shared with multiple other categories.
\end{enumerate}

The rational behind the definition of these three irregularities is as follows: very small or very large categories are likely too refined or too crude to adequately represent a scientific category. Large intersection between categories calls into question the necessity of accommodating at least one of these categories. Coverage may indicate a multi-disciplinary category, but, if the minimal coverage requires a considerable number of categories, it could also indicate that the covered category is ill-defined.   

When considering more than a single subject classification system, we further define three types of inter-system discrepancies:
\begin{enumerate}
    \item \textit{Different number of categories per journal}. 
    Journals in one system are classified to a significantly different number of categories in the other.  
    \item \textit{Similarity}: Subject categories in one system which exhibit a small (or empty) intersection with all categories in the second system.
     \item \textit{Coverage}: Categories in one system which can not be minimally covered by a very limited number of categories from the second system.
\end{enumerate}

The rational behind the definition of these three discrepancies as follows: A significantly different number of categories assigned to the same journal in each system suggests that one system is more lenient in its classification than the other. Small intersection could imply that the category in one system has no corresponding category in the other system. Finally, a category in one system which cannot be minimally covered by a very limited number of categories in the other system, is considered to be unsatisfactorily represented in the other. 

For the readers' convenience, relevant mathematical definitions and background on logical set theory is provided in Section \ref{subsec:math_def}}.

\SA{The journal subject classification systems of WoS and Scopus are widely used in practice and thus play a central role in the academic community \citep{pranckute2021web}.} 
As such, in this work, we focus on their \SA{two} journal subject classification systems and examine the \textit{intra-system} and \textit{inter-system}  \SA{irregularities and discrepancies as defined above}.
\SA{We speculate that these irregularities will prove prevalent in both systems and that significant discrepancies will be encountered across the two systems. }

%% file: Background.tex
\section{Background and Related Work}
\label{sec:background}

Our work focuses on WoS and Scopus which are the two most influential and most researched scholarly indexing services. Both index various types of source titles such as journals, conference proceedings and books. 

Starting with WoS, it has been traditionally considered a more reliable source for bibliometric analysis and extensive research has been conducted focusing on this index. 
Research analysing its journal subject classification system focused on the mapping of science and clustering based on its journal subject category classification system \citep{leydesdorff2013global, zhang2010subject}.
Other studies identified some of the problems associated \SA{with} this classification system. 
\citet{leydesdorff2016operationalization} showed that WoS subject categories are insufficient for performing bibliometric normalization due to \say{indexer effects}. They focused on the two fields- \say{Library and Information Science}, which has a WoS subject category and \say{Science and Technology Studies} which does not, and performed a mapping of citation behavior for journals in these fields. Their results showed that normalization using these categories might seriously harm the quality of the evaluation. 
\citet{haustein2012multidimensional} identified that WoS subject classification is controversial and problematic especially with regards to interdisciplinary fields due to the pigeonholing process taken when performing the classification. They claim that an alternative system to WoS subject classification is needed. In line with this recommendation, additional research by \citet{perianes2017comparison, shu2019comparing} compared WoS journal level classification system with publication level classification systems. They concluded that publication level classification systems constitute a credible alternative to WoS classification system. 
Following these studies, \citet{milojevic2020practical} presented a method for reclassification of WoS indexed articles into existing WoS categories as well as into 14 broad areas, based on the article references. 


Turning to Scopus, which is a more recent indexing system, little research was done on its journal subject classification system.
An early study by \citet{de2007coverage} focused on Scopus journal subject distribution, geographical distribution, language of publication among other measures. Their analyses shows that Scopus  has  quite  homogeneous  global  representation in nearly all areas except the Arts and Humanities. This study was conducted only 3 years after Scopus started indexing journals. A recent longitudinal analysis by \citet{bordignon2019tracking} observed the changes in number of categories per journal and number of journals per category. They showed an increase on average in both aspects and concluded that newly added sources have  been assigned to more fields and sub-fields on average than those indexed before the time period examined.
Their findings corroborate those found in \citet{wang2016large} which observed that Scopus journals are assigned to a large number of categories. Their analysis further identified some issues related to category naming including near identical names and categories which are labelled as \say{Miscellaneous}.
In \citet{lazic2017reliability}, the authors compared Scopus subject classification with the official classification of social sciences in Croatia and found a significant difference in the classifications. Their results showed Scopus mis-classifies  journals to social sciences subject categories despite them publishing almost exclusively works related to natural sciences or biomedicine.

Many studies have compared the two systems along with other indexing databases. The main focus of these studies were the accuracy of these databases. This accuracy was measured using the rankings both systems induced by ordering the retrieved publications in decreasing order of the number of citations \citep{bar2007some} or the citation links completeness and accuracy \citep{visser2021large, franceschini2016empirical}. These studies identified that both systems suffer from incompleteness and inaccuracy of citation links and incorrect transcription of author names and/or title. The work by \citet{meho2009assessing} compared Scopus and WoS using citations behavior focusing on \say{Information Science} researchers. Their findings show that when the analysis was based on small entities, such as journals and institutions, the scholarly impact measure produced by the two systems vary significantly, while analysis based on larger entities such as countries and research domains produced similar scholarly impact measure. They claimed that the need to use one or both indexing services will vary among research domains when used for assessing research impact.

Several studies analysed author related metrics generated from citations in these systems. In \citet{bar2008h} WoS, Scopus and Google Scholar (GS) were compared in terms of the h-index for a specific set of researchers. Their findings show that, except for a few cases, the differences in the h-indices between WoS and Scopus are not significant, but the differences between GS and the two other systems are much more considerable. \citet{harzing2016google} performed a longitudinal cross-disciplinary comparison of the WoS, Scopus and GS. Their results show that using the h-index with WoS as a data source, in the Life Sciences and Sciences was on average nearly eight times higher than in the Humanities. 

Other studies compared these systems in respect to the coverage and distribution of journals and publications.
These studies show that the difference in journal coverage between Scopus and WoS has grown over time and that differences in coverage resulted in variations in research output volumes, rank and global share of different countries \citep{singh2021journal, jacso2005we, mongeon2016journal}. In \citet{mongeon2016journal} the authors also observed that there is an over representation of certain countries and languages both in WoS and Scopus journal coverage. In addition, they show that WoS and Scopus journal coverage differ the most in the Natural Science and Engineering and in the Arts and Humanities fields. \citet{bartol2014assessment} showed that Scopus provides more records and more citations per record and, when focusing on disciplines, Scopus showed better coverage than WoS in Agriculture, Medical, and Natural Sciences and most noticeably in Engineering \& Technology.

Turning to the comparison of the journal subject category classification methodology used by  both systems, \citet{wang2016large} performed a detailed comparison of the classification systems of WoS and Scopus based on citation relations, where they measured the \say{connectedness} of a journal in respect to its assigned category and to other categories. They observed that, on average, journals have significantly more categories assignments in Scopus than in WoS. Furthermore, in Scopus journals are assigned to categories with which they are only weakly connected much more frequently than in WoS. They conclude that WoS and especially Scopus tend to be too lenient in assigning journals to categories.
\SA{A recent study by \citet{singh2020revisiting} analysed the accuracy of the subject classification systems of WoS, Scopus and Dimensions, by employing a user study where annotators ranked the classification accuracy of a large set of paper within each system. Their findings demonstrated that currently WoS is the most accurate classification system of the three and that the article level classification approach used in Dimensions may not necessarily outperform journal level classification as is done in Wos and Scopus.

Prior work has examined different aspects of WoS and Scopus indexing services 
(most recently \citet{martin2021google, singh2021journal, visser2021large, pranckute2021web}), separately and/or combined. However, to the best of our knowledge, logical set theory has yet to be applied in the investigation of journal subject classification systems. As such, our work complements on existing literature in this realm which has predominantly relied on citation analysis. }

\SA{We are not the first to adopt set theory-based analysis in bibliometrics in general. \citet{rodriguez2014evolutionary} performed graph based analysis where categories are modeled as vertices and edges represent the shared journals. 
\citet{subochev2018ranking} proposed ranking journals using methods from social choice and set theories to define aggregation methods of existing metrics.
Fuzzy sets theory, a variant of the classic set theory, was applied to 
bibliometric analysis in order to capture the \say{fuzzy} nature of field delineation \citep{bensman2001bradford, egghe2002proposal}.  In a study related to ours, \citet{rons2012partition}  proposed a normalization method based on partitioning WoS categories according on their intersecting journals. Their focus was on the adaptation of standard field normalization techniques based on these partitions. 
To our knowledge, our study is the first to examine irregularities and discrepancies in journals subject category classification systems using this approach.}



\subsection*{Mathematical Definitions}
\label{subsec:math_def}

Our study leverages aspects of the logical set theory \citep{cantor1874ueber}. Set theory is a branch of mathematical logic that studies the characteristics and relations between collections of objects. In this work, we consider sets of journals and, separately, sets of categories as needed. 

Let $A$ and $B$ be sets. 
We use the following operations  \citep{kolmogorov1975introductory}: 
\begin{itemize}
    \item Power - the number of members in A, denoted $|A|$.
    \item Union - the set of all members of A, B, or both, denoted $A \cup B$.
    \item Subset and Superset - set $A$ is a subset of set $B$ if all members of $A$ are also members of $B$, denoted $A \subseteq B$. if $A \subseteq B$ than $B$ is a superset of $A$.
    \item Equivalence - if   $A \subseteq B$ and $B \subseteq A$ then $A$ is equivalent to $B$, denoted $A=B$.
    \item Intersection - the set all members of $A$ which are also members of $B$, denoted $A \cap B$.
    \item Cover - a  collection of sets ($B,C,\cdots)$ excluding $A$ is said to cover $A$ if all members of $A$ are members of $B\cup C\cup\cdots$. A \say{minimal cover} of $A$ is the smallest number of sets needed to cover $A$.
\end{itemize}

%% file: Data_collection.tex
\section{Data}
\label{sec:data_collection}

Our study focuses on WoS and Scopus indexing systems.
From each of these systems we downloaded the complete set of the indexed journals, and their associated categories. Overall, 21,424 journals were extracted from WoS and 40,804 journals were extracted from Scopus. We excluded journals which were showing as \say{Discontinued} or \say{Inactive} in Scopus (WoS does not contain this data). The number of journals indexed in Scopus, which remained after this cleanup, was 25,751.

Since the two systems do not cover the same set of journals and one of the aims of our study is to perform a comparative analysis between the two, we focus only on journals which are indexed in both systems.
To identify these journals, journals from both systems were matched primarily based on their ISSN. In cases where the ISSN did not provide a match, we used the journals' e-ISSN as a secondary matching criteria\footnote{In order to perform these matches, cleanup was performed by removing the dash symbol which appeared in some of the identifiers as well as any leading zeros.}. Finally, for the very few cases not matched by ISSN or e-ISSN, we used the name as a matching criteria. The name matching was done as case insensitive exact matching. 
The final set of journals for our analysis comprised of 13,247 journals with 254 categories in WoS and 327 categories in Scopus\footnote{The category \say{Reviews and References (medical)} was excluded as it did not include any journals from our analysis.}. Overall, 8,177 journals from WoS and 12,504 journals from Scopus which were not matched by these identifiers were removed from further consideration.

All data and code is available in GitHub under
\textcolor{blue}{\href{https://github.com/shirAviv/journals_categories} {Journals Subject Classification}}.

\section{Methods}
\label{sec:Methods}

\SA{In order to identify intra-system irregularities and inter-system discrepancies, as defined in Section \ref{sec:intro}, we adopt a logical set theory approach. Specifically, we consider each subject category as a set of all the journals assigned to it.

Accordingly, the \textit{size} of the category is defined as the power of the set. We consider a category of less than 10 journals (less than 0.075\% of all journals) to be a very small one whereas one containing more than 300 journals (more than 2.5\% of all journals) to be a very big one. 
We measure the \textit{similarity} between pairs of categories as the size of the intersection divided by by the size of the smaller category of the two. We consider a pair of categories to be similar if their similarity measure exceeds 80\% and highly similar if their similarity measure exceeds 95\%. Categories with a 100\% similarity may either be equivalent or have a subset-superset relation as defined in Section \ref{subsec:math_def}. 
We compute the \textit{coverage} of each category using a standard minimal coverage algorithm. We consider both complete coverage (i.e., all journals are covered) as well as nearly-complete coverage (i.e., more than 90\% of all journals are covered). 

A Wilcoxon Signed Ranks Test \citep{wilcoxon1992individual} is applied to examine whether a significantly different number of categories is assigned to each journal across the two systems. 
}

%% file: Results_intra.tex
\subsection{Descriptive statistics}
\begin{figure}[ht]
    \subfloat[WoS]{
    \includegraphics[width=0.5\textwidth]{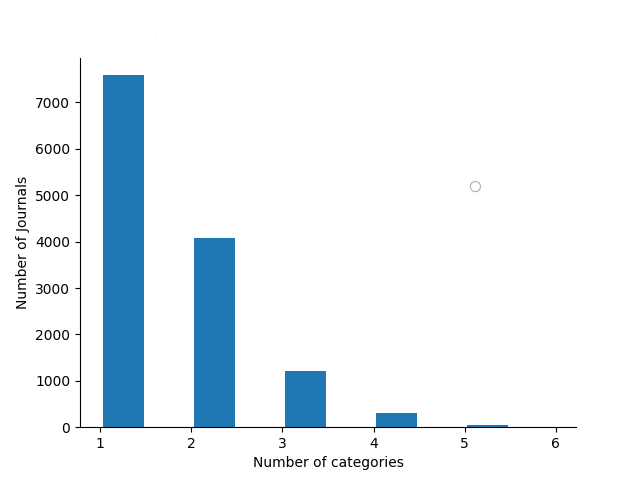}
    \label{fig:num_cats_per_journal_wos}}
~~~
    \subfloat[Scopus]{
    \includegraphics[width=0.5\textwidth]{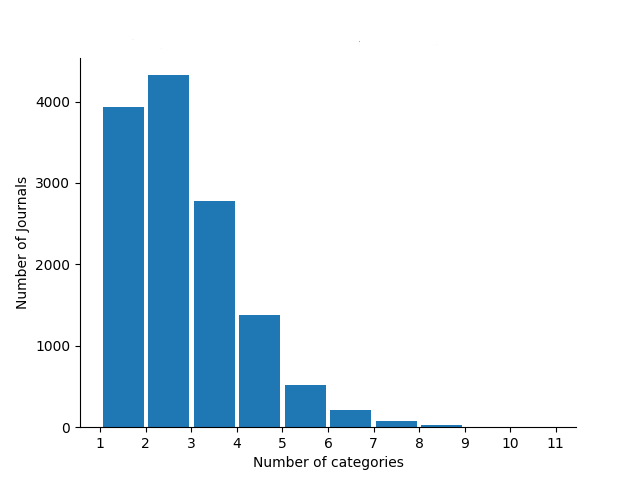}
    \label{fig:num_cats_per_journal_scopus}}
    \caption{Number of categories assigned to each journal.}
    \label{fig:num_cats_per_journal}
\end{figure}

\SA{In order to get an initial understanding of the data, we display the descriptive statistics for WoS and Scopus journals and subject categories in Table \ref{tbl:general_stats} and Figures \ref{fig:num_cats_per_journal_wos}, \ref{fig:num_cats_per_journal_scopus}.
Note that in Table \ref{tbl:general_stats} the mean, SD and median (bottom 6 rows) are calculated using only the journals and categories under analysis. The figures display the number of categories any single journal is classified to. It can be seen that, in WoS, the number of categories each journal is classified to decreases quickly, with most journals being classified to a single category and the highest number of classifications for a single journal is 6. In Scopus, however, more than 4,000 journals are classified to 2 categories (approximately a third of all journals under analysis), with one journal being classified to 11 different categories. \SA{These results align with prior works in the field (e.g \citet{wang2016large, pranckute2021web})}}

\begin{table}[ht]
    \centering
    \begin{tabular}{c|c|c}
        Descriptive Statistics & WoS & Scopus \\
        \toprule
        Number of categories & 254 & 327 \\ 
        Total number of journals & 21,424 & 40,804 \\
        Number of journals analysed & 13,247 & 13,247 \\
        \midrule
        Mean number of journals per category & 83.67 & 94.28 \\
        SD number of journals per category & 68.7 & 107.7 \\
        Median number journals per category & 63 & 67 \\
        \midrule
        Mean number of subject categories assigned to each journal & 1.58 & 2.34 \\
        SD number of subject categories assigned to each journal & 0.78 & 1.28 \\
        Median number of subject categories assigned to each journal & 1 & 2 \\
        
    \end{tabular}
    \caption{Descriptive statistics.}
    \label{tbl:general_stats}
\end{table}

 
\newpage
\subsection{Intra-System irregularities}
\label{subsec:intra}
\input{small_large_cats_table}

\begin{figure}[ht]
    \subfloat[WoS]{
    \includegraphics[width=0.5\textwidth]{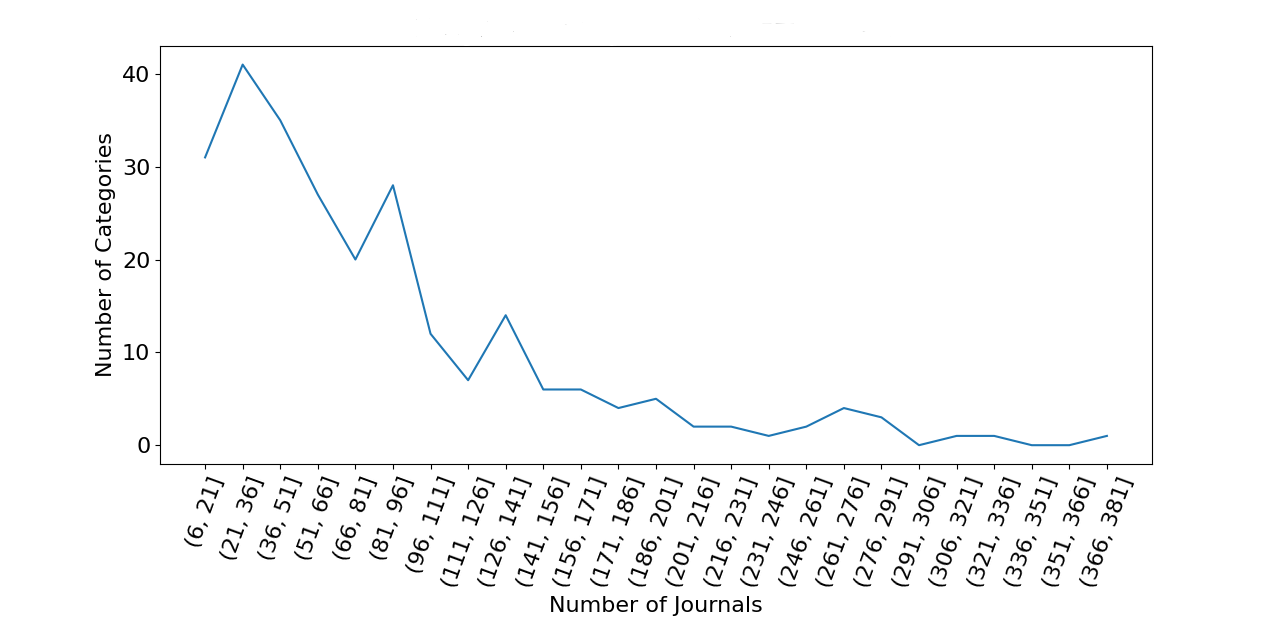}
    \label{fig:num_journals_per_cat_wos}}
    \subfloat[Scopus]{
    \includegraphics[width=0.5\textwidth]{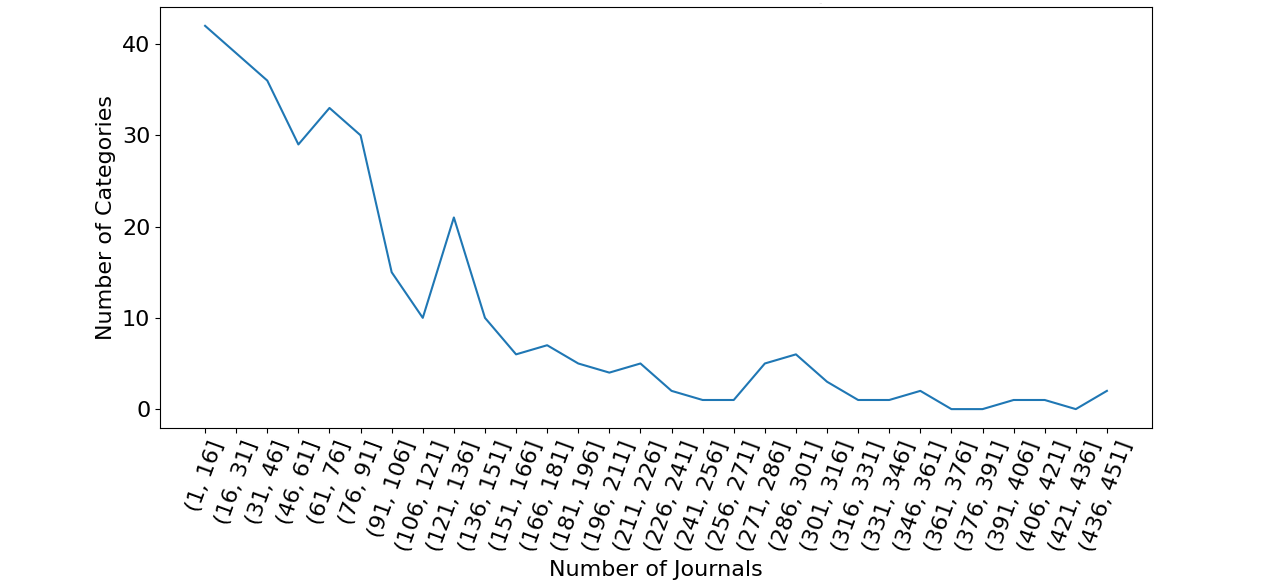}
    \label{fig:num_journals_per_cat_scopus}}
    \caption{Number of journals contained in each category. X-Axis is the number of journals (in bins of 15), Y-Axis is the number of categories containing the associated number of journals.}
    \label{fig:num_journals_per_cat}
\end{figure}

\SA{\subsubsection*{Size.}}
\label{subsub:Size}
Tables \ref{tbl:small_cats} and \ref{tbl:large_cats} display \SA{very small and very large} categories\SA{, respectively.} Starting with \SA{very small} categories, no category in WoS has less than 6 journals and only 3 categories are \SA{very small}. Scopus, on the other hand, has 8 categories containing only a single or two journals and 30 categories are \SA{very small}. Considering \SA{very large} categories, again, \SA{WoS has only 3 such categories while Scopus has 15 such categories} with the largest one containing 1139 journals, almost 10\%  of all analysed journals. 
Figure \ref{fig:num_journals_per_cat} displays the distribution of journals within categories for WoS and Scopus. As can be seen, the number of categories containing a large number of journals quickly decreases in both systems. It is also noticeable, both from Table \ref{tbl:small_cats} and from Figure \ref{fig:num_journals_per_cat}, that \SA{Scopus has a significantly larger number of both small and large categories than WoS.} This may be due to its higher number of total categories. These results seem to align with the statistics displayed in Table \ref{tbl:general_stats}, specifically, the large SD in Scopus.

\begin{figure}
   \subfloat[WoS]{
    \includegraphics[width=0.5\textwidth]{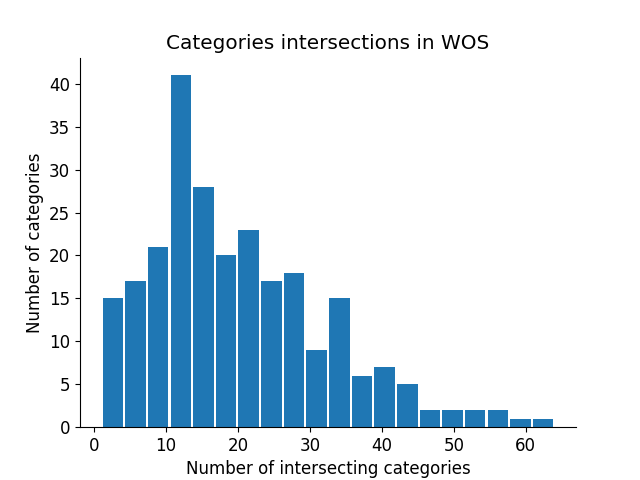}
    \label{fig:num_intersect_cats_per_cat-wos}}
~~~
    \subfloat[Scopus]{
    \includegraphics[width=0.5\textwidth]{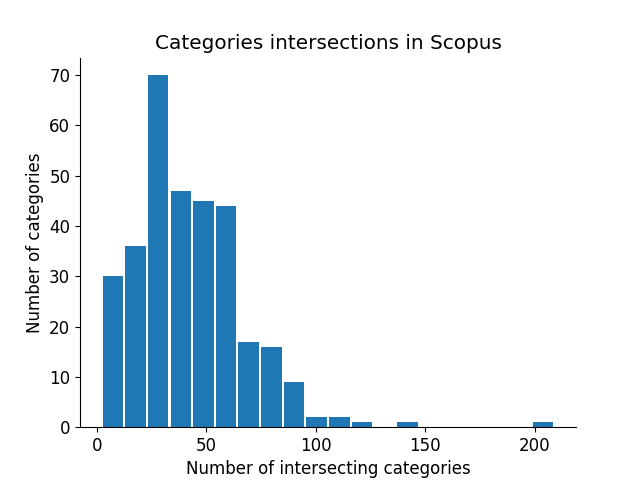}
    \label{fig:num_intersect_cats_per_cat-scopus}}
    
    \caption{Number of intersecting categories per category.}
    \label{fig:num_intersect_cats_per_cat}
\end{figure}

\newpage
\SA{\subsubsection*{Similarity.}} 
\label{subsub:intra_similarity}
Figures \ref{fig:num_intersect_cats_per_cat-wos} and \ref{fig:num_intersect_cats_per_cat-scopus} show the distribution of intersections of categories. While both WoS and Scopus show a similar distribution pattern, overall, Scopus categories intersect with many more categories. In WoS, no category intersects with more than 60 categories while, in Scopus, 55 categories have an intersection with over 60 other categories each. Moreover, 7 categories have intersections with over 100 categories and  the \say{General Medicine} category has intersections with over 200 categories. 

\newpage
The number of categories with an intersection with at least one other category in WoS is 252 - 99.2\% of all WoS categories.
Specifically, only 2 categories have no intersection with any other category. These are:
\begin{itemize}
    \item \say{Dance}, number of journals: 8
    \item \say{Literature, African, Australian, Canadian}, number of journals: 6
\end{itemize}
Note that these are 2 of the only 3 very small categories found in WoS (see Table \ref{tbl:small_cats}).
In Scopus, the number of categories with an intersection with at least one other category is 325 - 99.4\% of all Scopus categories.
Specifically, only 2 categories in Scopus have no intersection with any other category. However, these two categories contain only a single journal:
\begin{itemize}
    \item \say{Dental Hygiene}, number of journals: 1
    \item \say{Nurse Assisting}, number of journals: 1
\end{itemize}
\SA{Presumably, the single \say{Dental Hygiene} journal could be classified under \say{Dentistry (miscellaneous)} and the single journal under \say{Nurse Assisting} could be classified under \say{Nursing (miscellaneous)} or \say{Critical Care Nursing}.}

\begin{figure}
    \centering
    \includegraphics[width=0.5\textwidth]{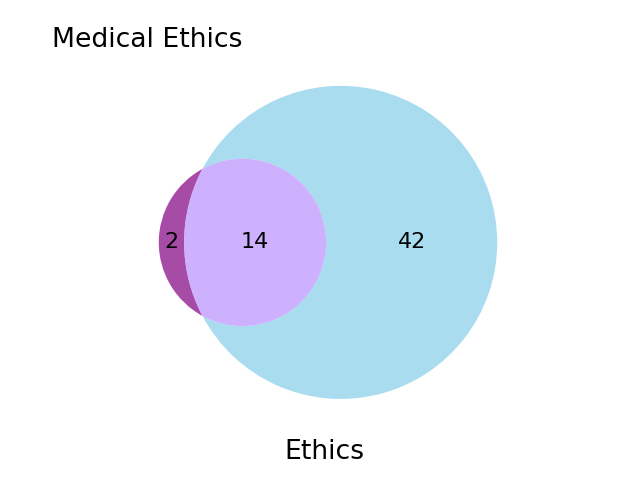}
    \caption{\SA{Similar WoS categories.}}
    \label{fig:wos_closest_intersect}
    
~~~~
   \subfloat[Highly similar.]{
    \includegraphics[width=0.33\textwidth]{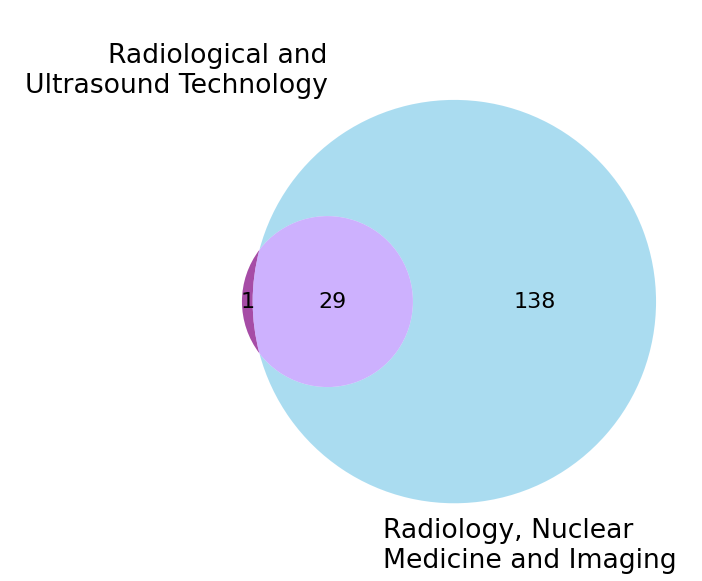}}
~~~~
    \subfloat[Similar.]{
    \includegraphics[width=0.33\textwidth]{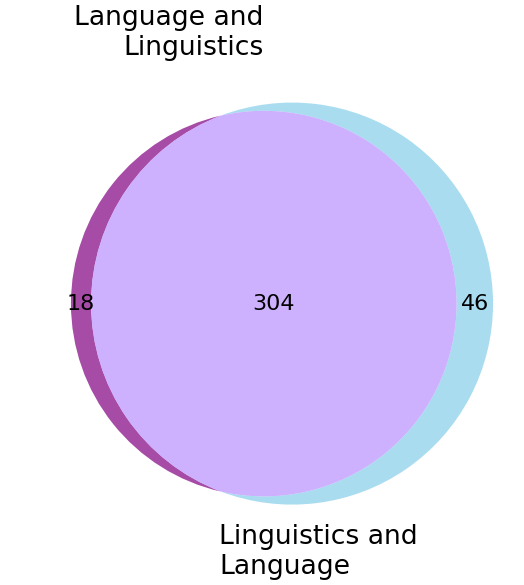}}
~~~~    
    \subfloat[Similar.]{
    \includegraphics[width=0.33\textwidth]{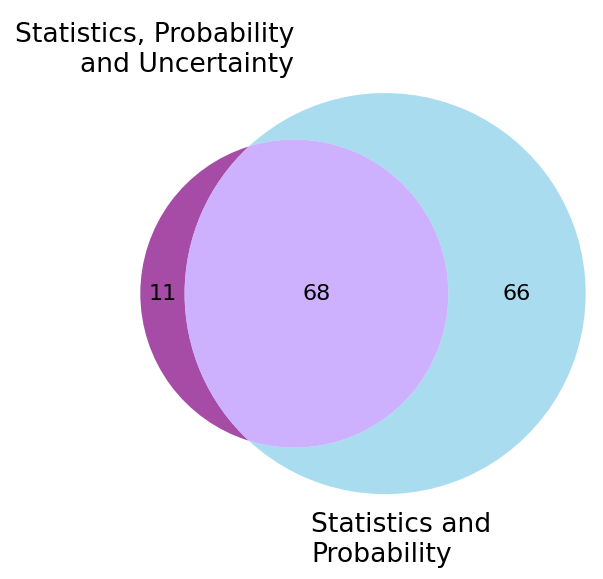}}
    
~~~
   \subfloat[Similar.]{
   \includegraphics[width=0.99\textwidth]{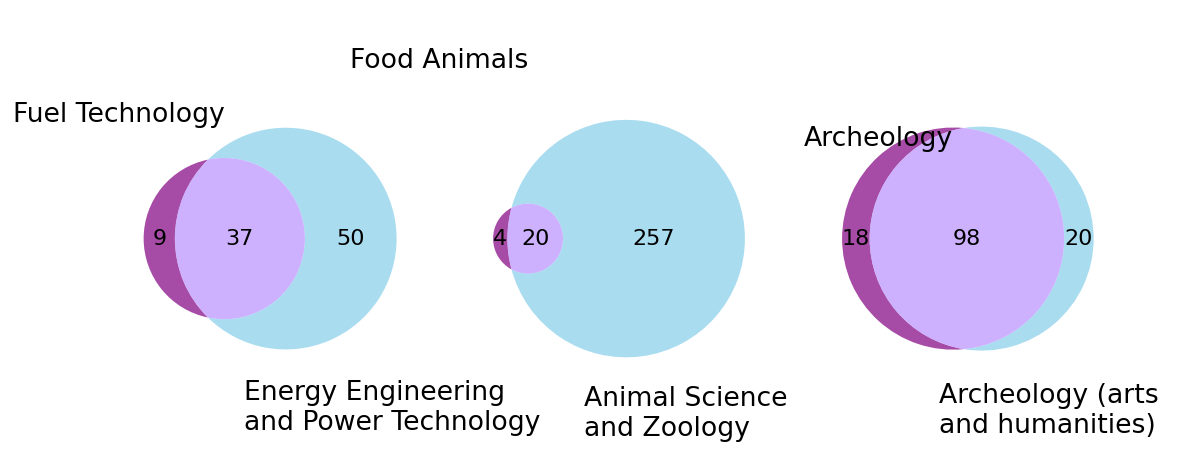}}
~~~
    \caption{Similar and highly similar Scopus categories.}
    \label{fig:closest_intersect_scopus}
\end{figure}

\begin{figure}[ht]
   \subfloat{
    \includegraphics[width=0.99\textwidth]{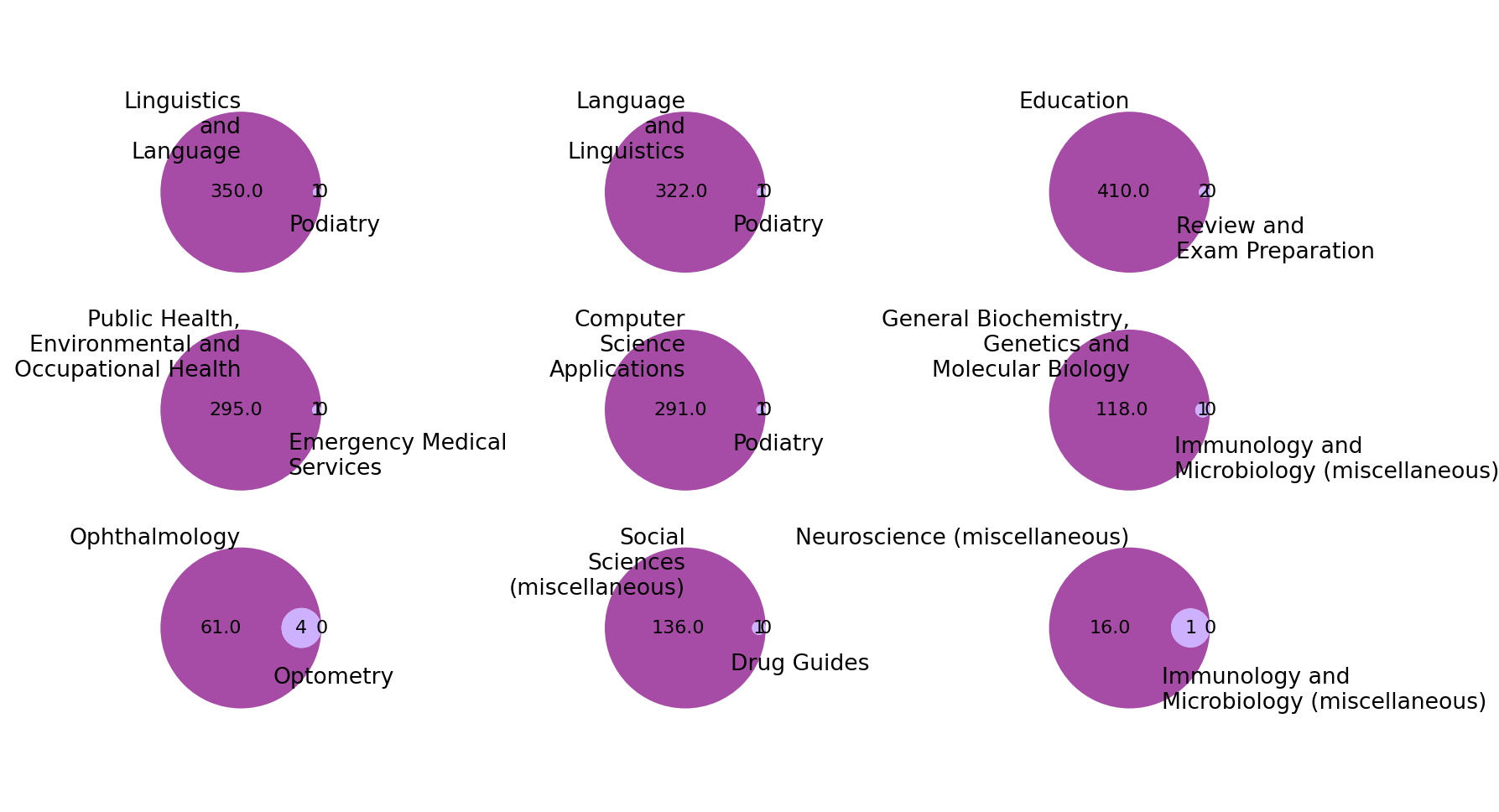}}
    ~~~
    \newline
    \subfloat{
    \includegraphics[width=0.65\textwidth]{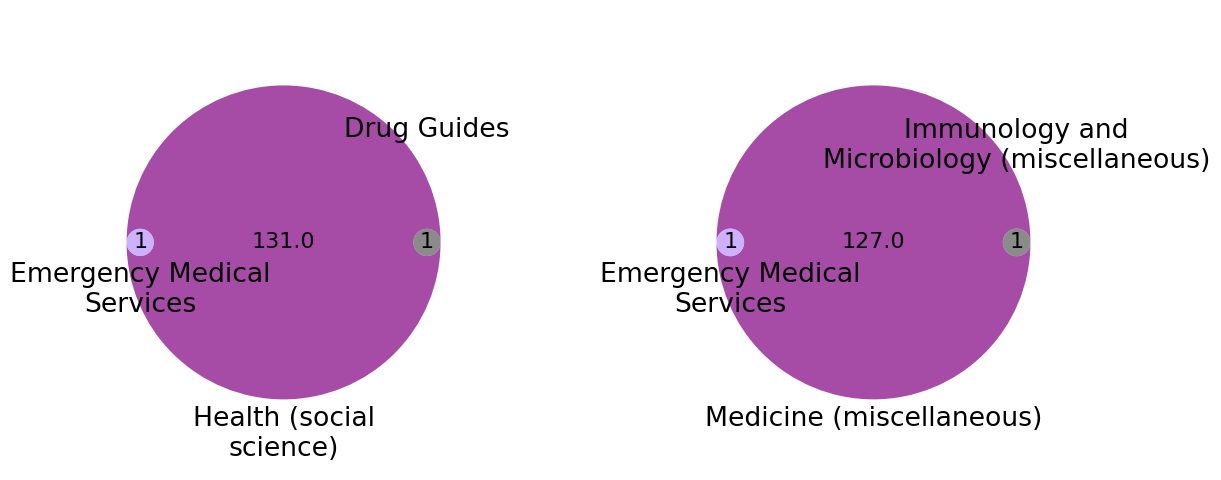}}
    
    \caption{Subsets of Scopus categories.}
    \label{fig:scopus_subsets}
\end{figure}

\SA{Focusing on similar and highly similar categories}, 
Figure \ref{fig:wos_closest_intersect} presents the single pair of \SA{similar} categories in WoS. Figure \ref{fig:closest_intersect_scopus} presents the 5 pairs of \SA{similar} categories and single pair of \SA{highly similar} categories in Scopus. 
\SA{No equivalent categories are found in either WoS or Scopus. No subset or superset categories are found in WoS. However, \SA{Scopus does have} 7 superset and 12 subset categories. Four of these are \SA{only} subsets of others, i.e., all journals in these categories are classified solely to them and to their superset categories}. Interestingly, these four categories are all single-journal categories as can be seen in Figure \ref{fig:scopus_subsets}. It can also be observed that the \say{Podiatry} category is a subset of three categories which, seemingly, are not related to it at all (\say{Language and Linguistics}, \say{Linguistics and Language} and \say{Computer Science Applications}).

\begin{figure}
\centering
    \includegraphics[width=\textwidth]{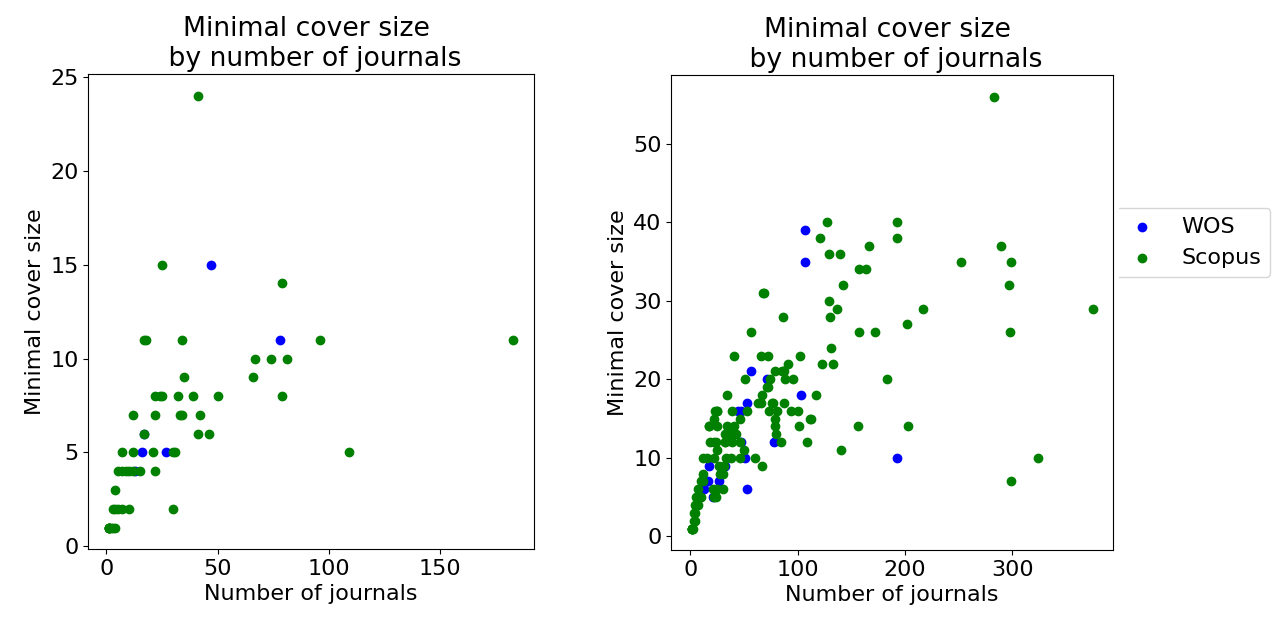}

    \caption{\SA{Minimal coverage in respect to number of journals. Complete (left) and nearly-complete (right) coverage, respectively.}}
    \label{fig:minimal_cover_intra}
\end{figure}

\newpage
\SA{\subsubsection*{Coverage.}}
\label{subsub:intra_minimal_cover}
In WoS, only 6 categories can be \SA{completely covered} ($\sim$2\%). All of these categories consist of between 13 and 78 journals in each category. The minimal number of categories needed to fully cover each of these six categories ranges from 4 to 15. Based on the very low number of categories which can be covered, we can deduce that all of the other categories in WoS ($\sim$98\%) have at least one journal which is classified solely to that specific category (and thus these categories can not be  \SA{completely covered}).
In Scopus, 56 categories ($\sim$17\%) can be \SA{completely} covered. These categories consist of between 1 and 183 journals in each category. The minimal number of categories needed to cover each of these categories ranges from 1 to 24 \SA{where 73\% of these categories require more than 3 categories for minimal complete coverage}. 
Recall that Scopus contains subset categories which naturally can be minimally covered by a single other category (i.e., the superset). 
Considering \SA{nearly-complete coverage}, in WoS, 14 additional categories can be covered ($\sim$8\%). The number of journals in each of these categories ranges from 13 to 193. The minimal number categories needed to cover these categories ranges from 5 to 39. In Scopus, 80 additional categories ($\sim$41\%) can be covered, with number of journals ranging from 1 to 375. The minimal number of categories needed to cover each of these  categories ranges from 1 to 56 \SA{and over 90\% of the categories require at least 4 categories to minimally cover them}. 
Figure \ref{fig:minimal_cover_intra} displays the minimal number of categories required for \SA{complete or nearly-complete coverage} in respect to the number of journals in each category.  

\SA{When considering nearly-complete coverage, a significantly larger number of categories can be covered, most prominently in Scopus. Most of these added categories contain less than 100 journals.} 
\newpage

%% file: small_large_cats_table.tex
\begin{table}
\centering

\begin{tabular}{llll}

\toprule
\multicolumn{2}{c}{Scopus} &
\multicolumn{2}{c}{Web Of Science} \\
\hline
Category &  Num Journals &     Category &    Num Journals  \\

\midrule
Dental Hygiene & 1 & Literature, African, Australian, Canadian          &            6 \\
Nurse Assisting & 1 & Andrology &   8 \\
Drug Guides & 1 &   Dance     & 8       \\
Emergency Medical Services & 1 &     &          \\
Podiatry & 1 &          &             \\
\multirow{2}{*}{Immunology and Microbiology}\\{ (miscellaneous)} & 1 &  &    \\
Pharmacology (nursing) & 2 &          &             \\
Review and Exam Preparation & 2 &          &             \\
Assessment and Diagnosis & 3 &          &             \\
Care Planning & 3 &          &             \\
Chiropractics  & 3 &          &             \\
Optometry   & 4 &          &             \\
\multirow{2}{*}{Pharmacology, Toxicology}\\ {and Pharmaceutics (misc)} & 4 & & \\
Complementary and Manual Therapy & 4 &  &       \\
Pharmacy   & 5 &          &             \\
Fundamentals and Skills & 5 &  & \\
General Health Professions & 6 &          &      \\
Equine   & 6 &          &             \\
Research and Theory & 6 &          &             \\
Chemical Health and Safety & 7 &          &      \\
Veterinary (miscellaneous) & 7 &          &       \\
Dentistry (miscellaneous) & 7 &          &       \\
Periodontics & 8 &          &             \\
Health Professions (miscellaneous) & 8 &    &    \\
Medical and Surgical Nursing & 9 &          &     \\
Pediatrics       & 9 &          &             \\
Critical Care Nursing    & 9 &          &             \\
Occupational Therapy     & 9 &          &             \\
Emergency Nursing    & 9 &          &             \\
Nursing (miscellaneous) & 9 &          &       \\
\bottomrule
\end{tabular}
\caption{Small categories in Web Of Science and Scopus, less than 10 journals in category.}
\label{tbl:small_cats} 
\end{table}

\begin{table}[t]
\centering
\begin{tabular}{llll}

\toprule
\multicolumn{2}{c}{Scopus} &
\multicolumn{2}{c}{Web Of Science} \\
\hline
Category &  Num Journals &     Category &    Num Journals  \\

\midrule
Molecular Biology & 300 & Materials Science, Multidisciplinary & 307 \\
Mechanical Engineering & 304 & Mathematics  & 322 \\
Biochemistry & 305 & Economics  &  372 \\
Condensed Matter Physics & 306 & & \\
Language and Linguistics & 322 & & \\
Economics and Econometrics & 340 & & \\
Linguistics and Language & 350 & & \\
Electrical and Electronic Engineering & 352 & & \\ 
Cultural Studies & 402 &   &   \\
Education & 410 &   &   \\
\multirow{2}{*}{Ecology, Evolution, Behavior}\\{ and Systematic}s & 442 &   &   \\
Literature and Literary Theory & 448 &   &   \\
History & 532 &   &   \\
Sociology and Political Science & 537 &   &   \\
General Medicine & 1139 &   &   \\
\bottomrule
\end{tabular}
\caption{Large categories in Web Of Science and Scopus, more than 300 journals in category.}
\label{tbl:large_cats} 
\end{table}

%% file: Results_inter.tex
\subsection{Inter-System discrepancies}

\subsubsection*{Different number of categories per journal.}
The number of categories assigned to a journal in Scopus is statistically significantly higher than number of categories assigned in WoS, $(Z=3979962.5, p<0.001)$.
On average, a journal is classified to 2.34 categories in Scopus, whereas it is classified to 1.58 categories in WoS. This finding may suggest that Scopus uses a more lenient classification of journals to subject categories than WoS does. If that is the case, one would expect to identify substantial similarities between the two systems. This is  examined next.

\subsubsection*{Similarity.}
Recall that, in this work, we consider only the journals indexed by both systems. Thus, all categories in one system have at least one intersection with at least one category in the other.

\begin{figure}[ht]
    \centering
    \includegraphics[width=0.95\textwidth]{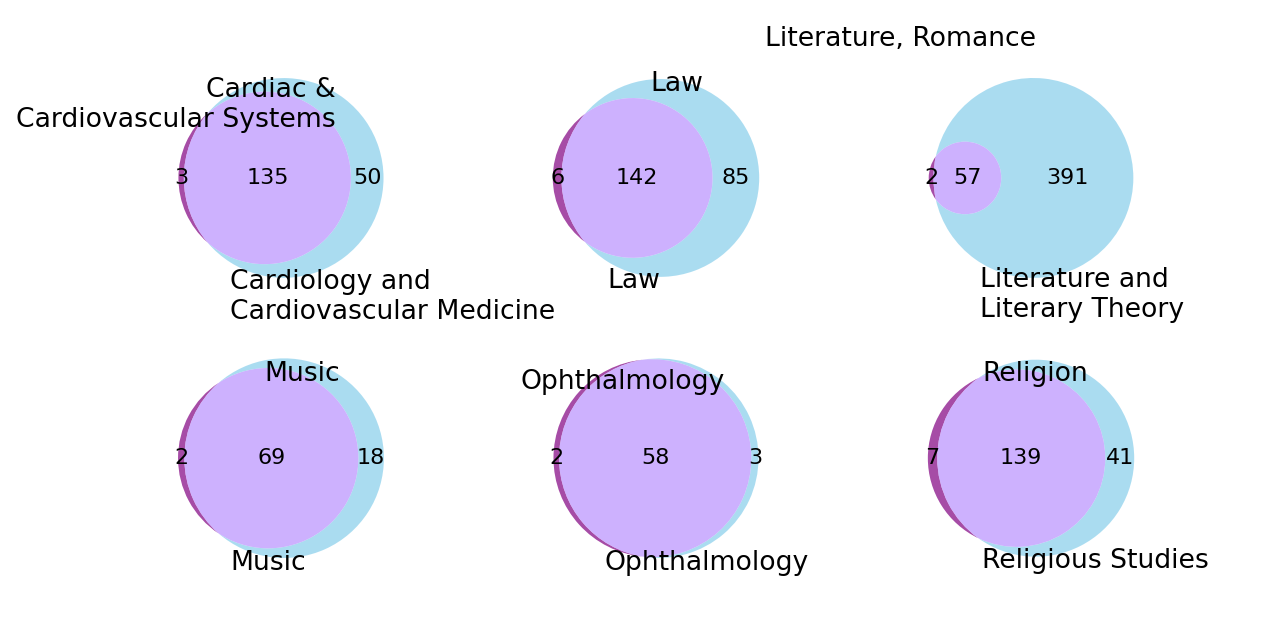}
    \caption{\SA{Highly similar WoS to Scopus categories.} Purple- WoS categories, Cyan - Scopus categories.}
    \label{fig:wos_scopus_intersect_95}
\end{figure}

\begin{figure}
    \centering
    \includegraphics[width=0.4\textwidth]{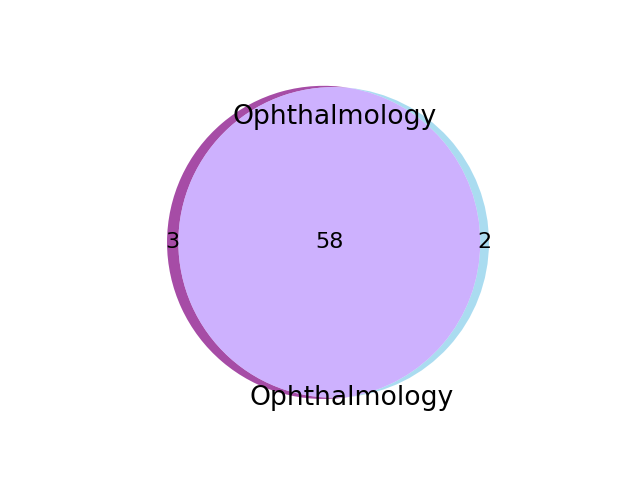}
    \caption{\SA{Highly similar Scopus to WoS categories.} Purple- Scopus category, Cyan - WoS category.}
    \label{fig:scopus_wos_intersect_95}
\end{figure}

Starting with WoS, \SA{nearly 30\% of its categories have a similar category in Scopus, and only 6 highly similar categories are found} (see Figure \ref{fig:wos_scopus_intersect_95} for the \SA{highly similar} categories). 
Turning to Scopus, \SA{only $\sim$20\% of its categories have a similar counterpart in WoS, and only a single highly similar category is found}  (see Figure \ref{fig:scopus_wos_intersect_95}). The names of these categories roughly match. For example, \say{Cardiac $\&$ Cardiovascular Systems} in WoS is \SA{highly similar} to  \say{Cardiology and Cardiovascular Medicine} in Scopus. \SA{The low number of similar categories may suggest that the 2 systems sparsely coincide in their category definition.}  

\begin{figure}[H]
    \centering
    \includegraphics[width=0.75\textwidth]{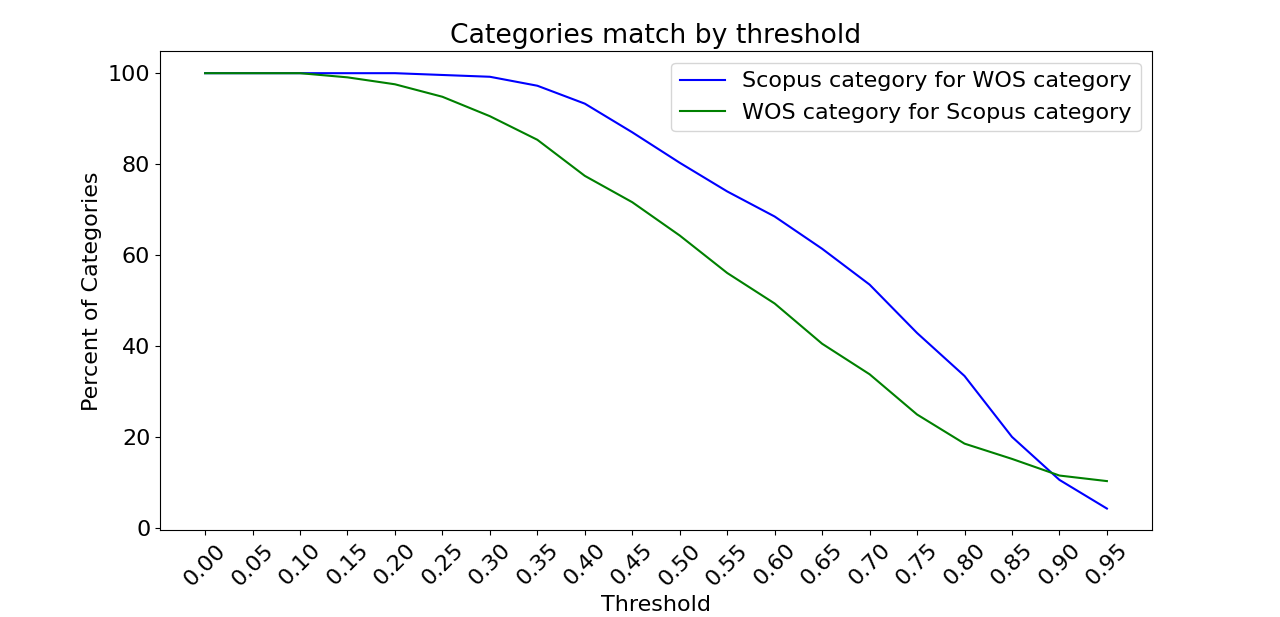}
    \caption{Match between WoS categories and Scopus categories based on percentage of shared journals in respect to total journals in the category.}
    \label{fig:threshold_match}
\end{figure}

Figure \ref{fig:threshold_match} displays the results of \SA{a more relaxed definition of similarity, i.e., less than 80\% shared journals.} It can be seen that for low percentage of shared journals, as can be expected, all categories in one system have \say{enough} shared journals with at least one category in the second system. \SA{However, as the required percentage of shared journals increases, the number of similar categories quickly declines.}
The fact that the blue line is above the green line at nearly all points in the plot indicates that \SA{WoS's categories have similar Scopus categories} for most examined shared journals percentages. 

\begin{figure}
    \centering
    \includegraphics[width=0.95\textwidth]{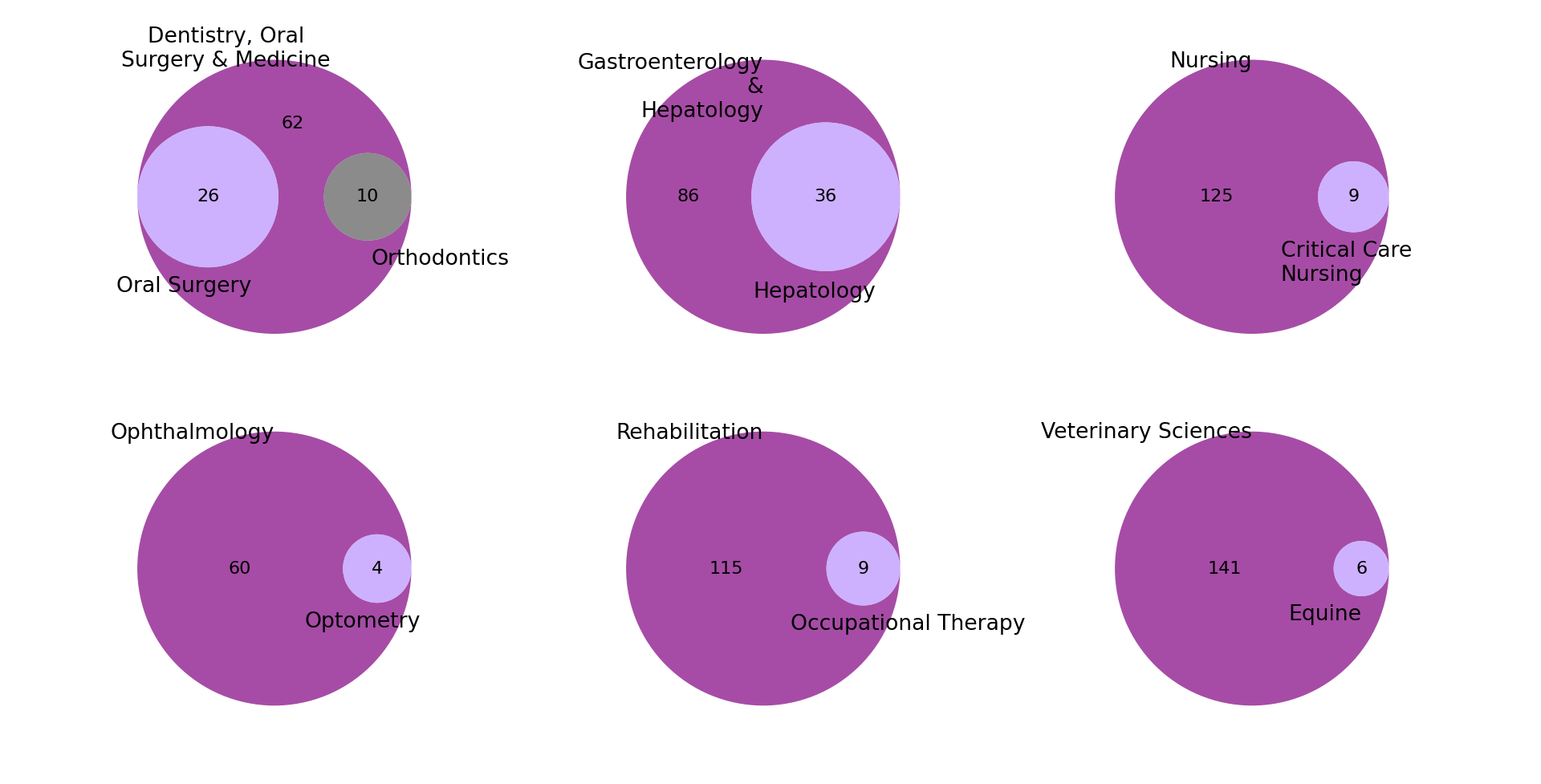}
    \caption{\SA{WoS categories which have Scopus subsets.} Dark purple - WoS categories, lilac and gray - Scopus categories.}
    \label{fig:wos_scopus_subsets}
\end{figure}
\begin{figure}
       \centering
     \includegraphics[width=0.95\textwidth]{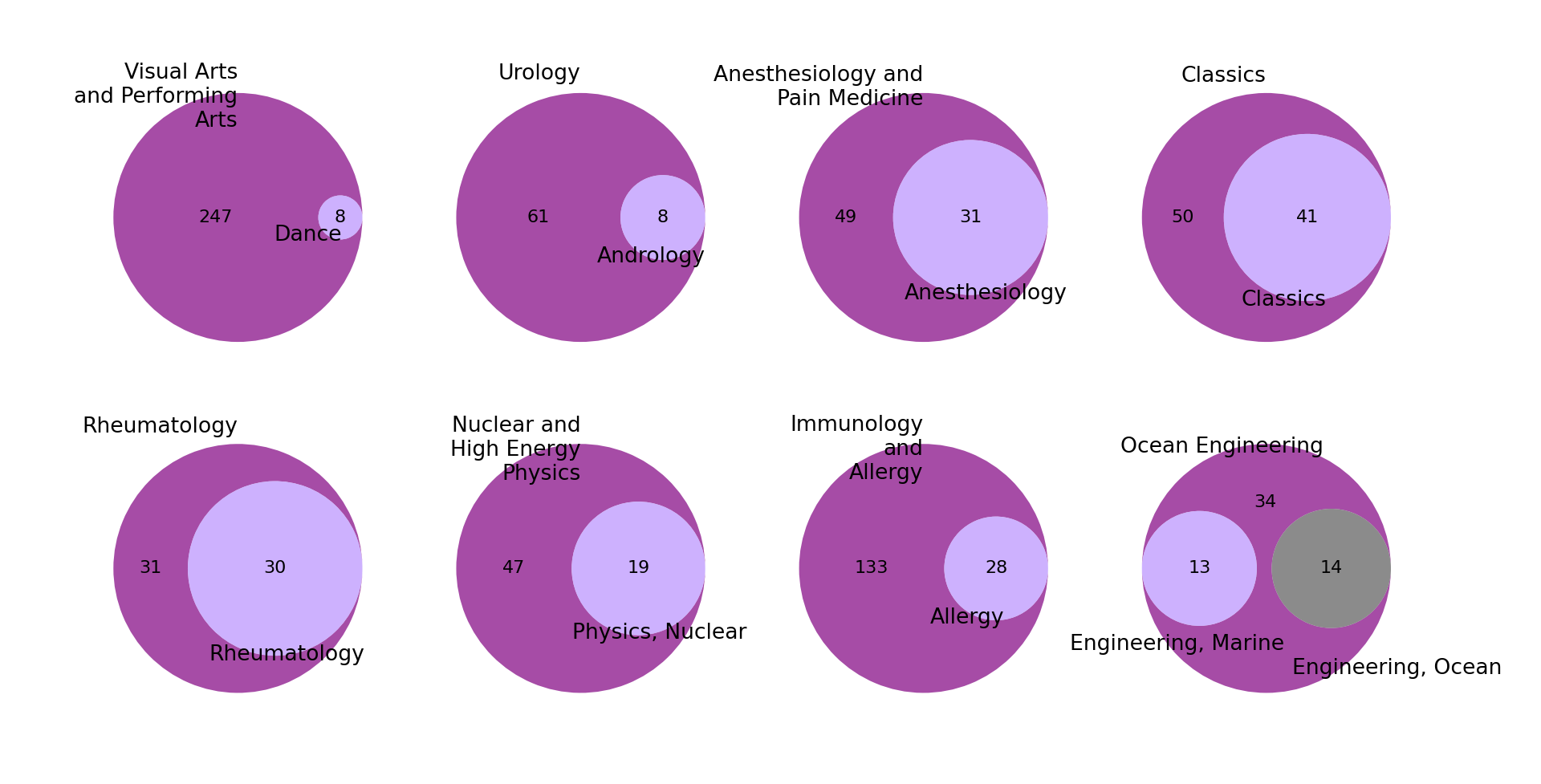}
    \caption{\SA{Scopus categories which have WoS subsets.} Dark purple - Scopus categories, lilac and gray - WoS categories.}
    \label{fig:scopus_wos_subsets}
\end{figure}

No equivalent categories between WoS and Scopus are found. 
However, 13 categories in WoS have subsets in Scopus and 9 categories in Scopus have subsets in WoS. Figures \ref{fig:wos_scopus_subsets} and \ref{fig:scopus_wos_subsets} display the categories with more than a single journal. The names of these categories are very indicative of their relation.
\SA{For example, WoS category \say{Dentistry, Oral Surgery $\&$ Medicine} is a superset of \say{Oral Surgery} and \say{Orthodontics}, two of Scopus's categories. }

\newpage
\SA{\subsubsection*{Coverage.}}
Starting with WoS, the minimal number of categories from Scopus required to \SA{completely} cover all of WoS's journals is 279 (85\% of the total Scopus categories). 
\SA{In order to achieve a nearly-complete coverage, 130 categories are needed (40\%)}.
Turning to Scopus, the minimal number of categories from WoS required to \SA{completely} cover Scopus's journals is 248 (98\% of WoS categories) 
\SA{In order to achieve a nearly-complete coverage, 140 categories are needed (55\%)}. Taken jointly,  these results indicate that the number of categories required to cover all journals in one system by the other is similar for both systems. Interestingly, in Scopus, \SA{50 categories are not needed for the complete coverage of WoS. If one were to remove these categories from Scopus, both systems were to have roughly the same number of categories.}  

\begin{figure}
   \subfloat[WoS]{
    \includegraphics[width=0.5\textwidth]{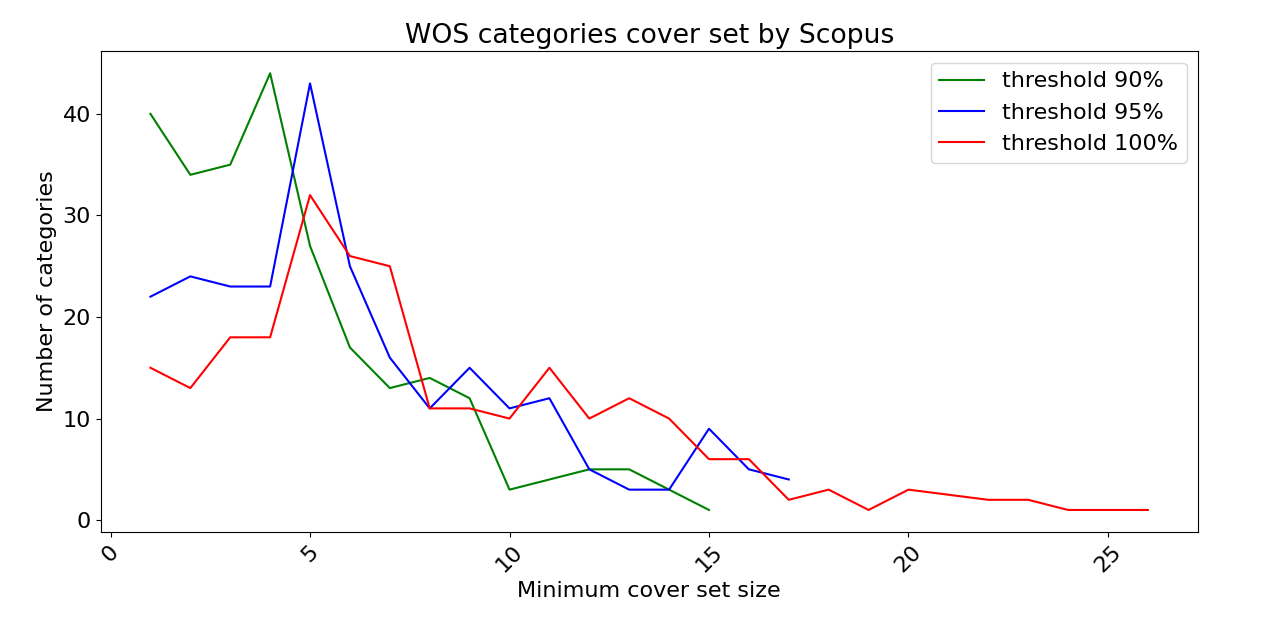}
    \label{fig:wos_scopus_min_cover_set}}
~~~
    \subfloat[Scopus]{
     \includegraphics[width=0.5\textwidth]{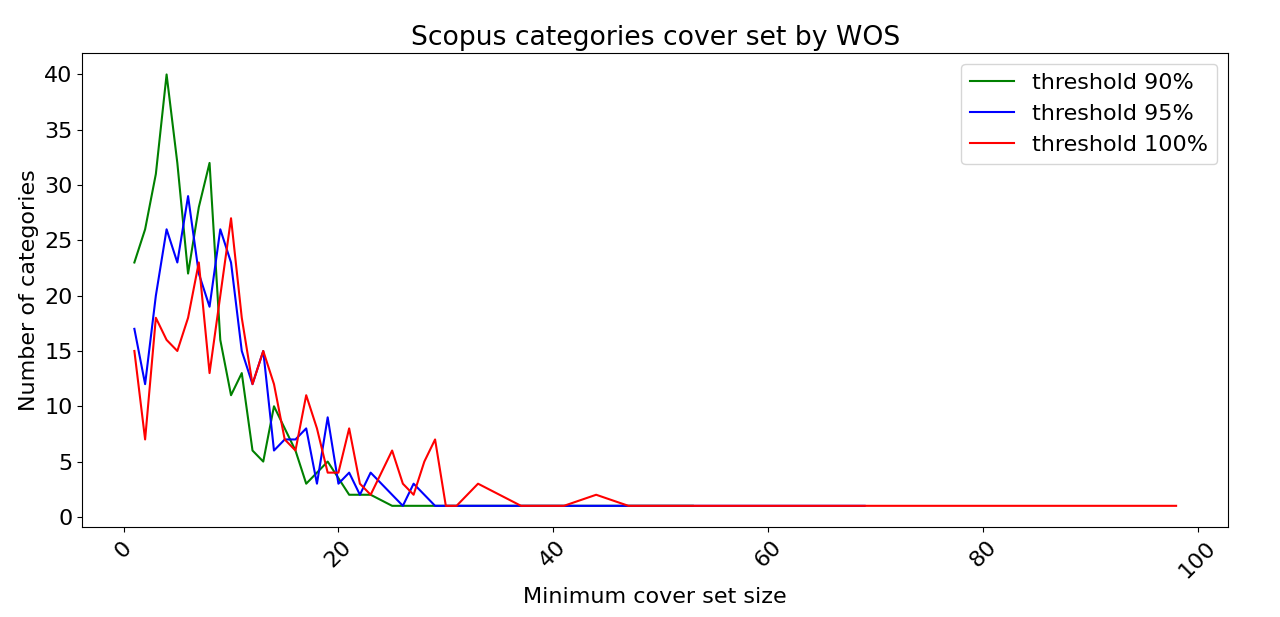}
    \label{fig:scopus_wos_min_cover_set}}
    
    \caption{\SA{Complete and nearly-complete} coverage.}
    \label{fig:min_cover_set_inter}
\end{figure}

\begin{figure}
   \subfloat[WoS]{
    \includegraphics[width=0.5\textwidth]{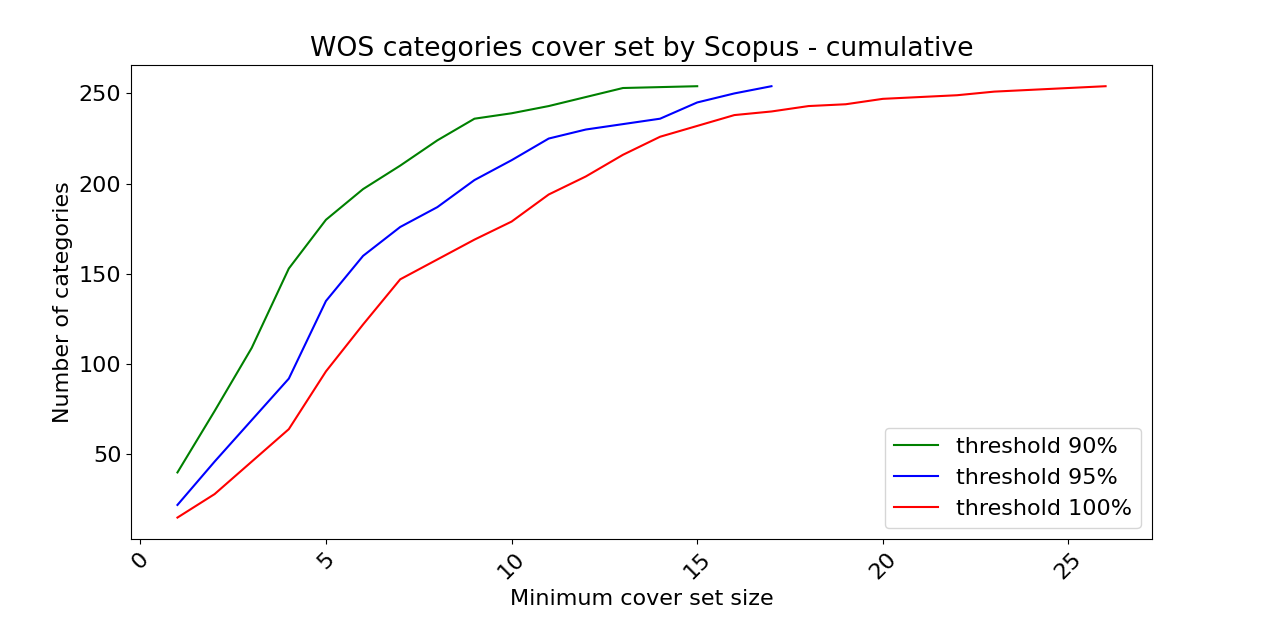}
    \label{fig:wos_scopus_min_cover_set_cumulative}}
~~~
    \subfloat[Scopus]{
     \includegraphics[width=0.5\textwidth]{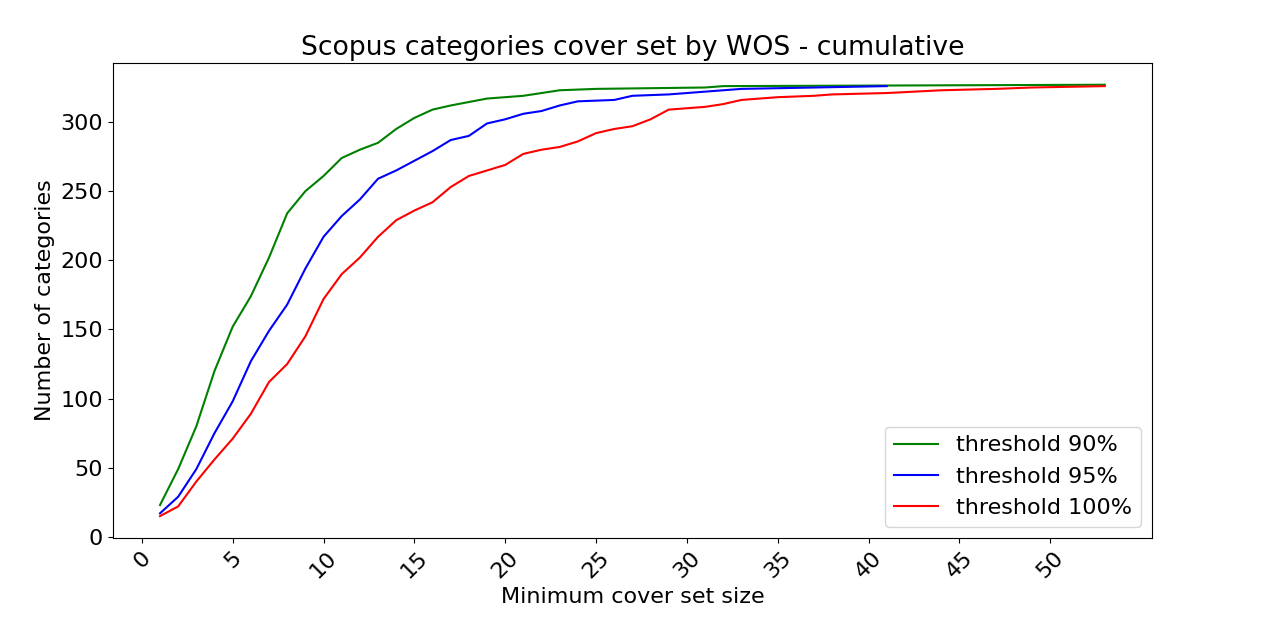}
    \label{fig:scopus_wos_min_cover_set_cumulative}}

    \caption{Cumulative \SA{complete and nearly-complete} coverage.}
    \label{fig:min_cover_set_inter_cumulative}

\end{figure}

The minimal number of categories from Scopus required to \SA{completely} cover each single WoS category ranges from 1 to 26. The minimal number of categories from WoS required to \SA{completely} cover each single Scopus category ranges from 1 to 98. 
Figure \ref{fig:min_cover_set_inter} \SA{displays the minimal number of categories in one system required to completely cover each categories in the second system}. Figure \ref{fig:min_cover_set_inter_cumulative} displays the cumulative minimal cover needed for each system.
As can be seen \SA{from both figures},
in WoS, \SA{only $\sim$$20$\% of the categories are can be completely covered by up to of 3 categories from Scopus. 
Even in the case of nearly-complete coverage, only $\sim$40\% of the categories can be minimally covered by up to 3 categories.} Scopus shows a similar trend as \SA{only $\sim12$\% of its categories can be completely covered and only $\sim$27\% of the categories can be nearly-completely covered by up to 3 categories from WoS.}
\SA{These findings suggest that the minimal number of categories required for coverage, either complete or nearly-complete, of any single category is at least 4 for the majority of categories. For example, for nearly-complete coverage, $\sim$$30\%$ and $\sim$$50\%$ of the categories in WoS and Scopus, respectively, require at least 5 categories.}

\begin{figure}
    \centering
    \includegraphics[width=0.7\textwidth,height=0.5\textwidth]{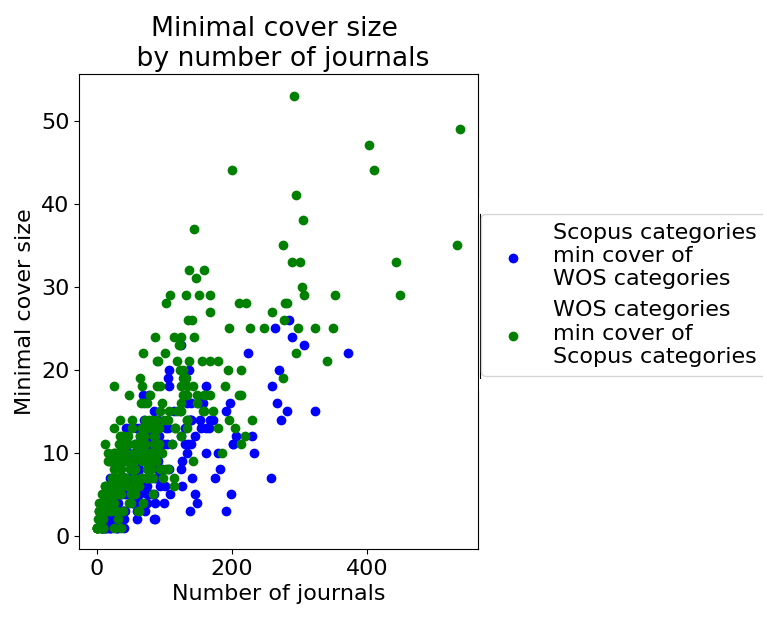}
    \caption{\SA{Complete Scopus coverage of WoS categories and complete WoS categories coverage of Scopus categories.} Outliers removed.}
    \label{fig:min_cover_set_no_ouliers}
\end{figure}
Figure \ref{fig:min_cover_set_no_ouliers} displays the relations between the size of the \SA{minimal complete coverage and the number of journals in each category.} It is noticeable that about $50\%$ of the categories are \SA{minimally} covered by 5 \SA{or more} categories from the second system. 
\SA{Our findings concerning coverage further strengthen the conclusion that categories in either system are inadequately represented in the other system, or not represented at all. }



%% file: conclusion.tex
\newpage
\section{Conclusions}
\label{sec:conclusion}

In this study we have examined the intra-system irregularities and inter-system discrepancies  the journal subject classification systems of WoS and Scopus. 

\SA{Overall, our results indicate that irregularities are common in both systems. Specifically, in both systems we identify unusually sized categories, high overlap and incohesiveness between categories. These are especially prominent in Scopus.
Discrepancies between the two systems are prevalent too. Specifically, journals are systematically classified to a different number of categories and most categories are not adequately represented in the other system with either one or a limited number of corresponding categories. 
As such, one system should not be considered as either a \say{finer grained version} or \say{close variant} of the other.}

\SA{Consequently, relying on either one of these systems may misguide their users, potentially leading to unwarranted outcomes and conclusions which may also prove inconsistent when  considering the other system. 
This may pose a significant concern when relying on these journal subject classification systems for scientometric research and practice. As such, users should consider alternatives to these two systems. 
Above all, our results advocate the standardization of journal subject classification.}

\SA{Irregularities in journal subject classification systems may also pose a significant concern regarding journal scientometric scores and rankings, which often rely on the journal's associated categories. In future work, we plan to examine whether the identified irregularities lead to potential distortions in scores and rankings. }
In addition, we wish to further expand our study to include additional indexing services such as Dimensions and Google Scholar.

\subsection*{\SA{Limitations}}
We recognize that this study is limited in several aspects.
First, our \SA{data pre-processing} could have overlooked journals with slightly different names or incorrect ones. Note that we have performed a manual analysis of some of the journals in order to mitigate this issue as much as possible.
In addition, our study focused only on journals indexed by both systems in order to identify discrepancies. As such, our findings are limited to the journals and categories under analysis. \SA{Last, our study has not considered or proposed a specific method for rectifying the identified irregularities and discrepancies. }